\documentclass[%
 reprint,
 amsmath,amssymb,
 longbibliography,
 prl
]{revtex4-2}

\usepackage[ruled]{algorithm2e}
\usepackage[noend]{algpseudocode}
\usepackage{type1cm} 
\usepackage{graphicx}
\usepackage{dcolumn}
\usepackage{bm}
\DeclareMathAlphabet{\mathpzc}{OT1}{pzc}{m}{it}

\begin{document}


\title{Correlation Effects and Concomitant Two-Orbital $s_\pm$-Wave Superconductivity in La$_3$Ni$_2$O$_7$ under High Pressure}

\author{Yi-Heng Tian}
\affiliation{Department of Physics, Renmin University of China, Beijing 100872, China}
\affiliation{Key Laboratory of Quantum State Construction and Manipulation (Ministry of Education), Renmin University of China, Beijing 100872, China}

\author{Yin Chen}
\affiliation{Department of Physics, Renmin University of China, Beijing 100872, China}
\affiliation{Key Laboratory of Quantum State Construction and Manipulation (Ministry of Education), Renmin University of China, Beijing 100872, China}

\author{Jia-Ming Wang}
\affiliation{Department of Physics, Renmin University of China, Beijing 100872, China}
\affiliation{Key Laboratory of Quantum State Construction and Manipulation (Ministry of Education), Renmin University of China, Beijing 100872, China}

\author{Rong-Qiang He}\email{rqhe@ruc.edu.cn}
\affiliation{Department of Physics, Renmin University of China, Beijing 100872, China}
\affiliation{Key Laboratory of Quantum State Construction and Manipulation (Ministry of Education), Renmin University of China, Beijing 100872, China}

\author{Zhong-Yi Lu}\email{zlu@ruc.edu.cn}
\affiliation{Department of Physics, Renmin University of China, Beijing 100872, China}
\affiliation{Key Laboratory of Quantum State Construction and Manipulation (Ministry of Education), Renmin University of China, Beijing 100872, China}

\date{\today}

\begin{abstract}
  Possible high-$T_c$ superconductivity (SC) has been found experimentally in the bilayer material La$_3$Ni$_2$O$_7$ under high pressure recently, in which the Ni-$3d_{3z^2-r^2}$ and $3d_{x^2-y^2}$ orbitals are expected to play a key role in the electronic structure and the SC. Here we study the two-orbital electron correlations and the nature of the SC using the bilayer two-orbital Hubbard model downfolded from the band structure of La3Ni2O7 in the framework of the dynamical mean-field theory. We find that each of the two orbitals forms $s_\pm$-wave SC pairing. Because of the nonlocal inter-orbital hoppings, the two-orbital SCs are concomitant and they transition to Mott insulating states simultaneously when tuning the system to half filling. The Hund's coupling induced local inter-orbital spin coupling enhances the electron correlations pronouncedly and is crucial to the SC.
\end{abstract}


\maketitle



Studying more than three decades, the superconducting (SC) mechanism of cuprates as the first class of unconventional high-temperature superconductors remains elusive \cite{Keimer2015nat}. While undoped cuprates are antiferromagnetic Mott insulators, in doped cuprates the SC takes place in the CuO$_2$ layer and shows $d_{x^2-y^2}$-wave symmetry. In-plane antiferromagnetic spin correlations are thought to be crucial for the Cooper pairing. Because of the large Coulomb interaction of the active Cu-$3d_{x^2-y^2}$ orbital, cuprates are strongly electronically correlated. Doped cuprates are thought as doped Mott insulators \cite{Mott1968rmp,Imada1998rmp,Lee2006rmp} and are usually described by a single-band Hubbard or $t$-$J$ model in theoretical studies. Besides cuprates, there are other SC strongly correlated electronic materials being discovered \cite{kamihara2008iron,Chubukov2012ARC,Li2019Nature}.

Recently, the bilayer Ruddlesden-Popper phased La$_3$Ni$_2$O$_7$ under high pressure was found to be another high-temperature superconductor \cite{Sun2023Nature} and quickly attracted a lot of attention \cite{Zhang2023Electronic,gu2023Effective,christiansson2023Correlated,lechermann2023Electronic,sakakibara2023Possible,yang2023Possible,shen2023Effective,shilenko2023Correlated,liu2023Electronic,wu2023Charge,cao2023Flat,chen2023Critical,liu2023The,zhang2023high,oh2023type,liao2023electron,lu2023interlayer,qu2023bilayer,zhang2023structural,yang2023minimal,jiang2023high,zhang2023trends,huang2023impurity}. Reminiscent of the CuO$_6$ octahedra in cuprates, the bilayer stacked NiO$_6$ octahedra, are the key structure for the SC. Like cuprates, the La$_3$Ni$_2$O$_7$ features a linear resistivity behavior in the high-temperature normal state, implying that its normal state is a strange metal and it may be a strongly correlated electronic material. As the density functional theory (DFT) based band structure calculations \cite{Sun2023Nature} show, the three Ni-$t_{2g}$ orbitals are fully filled while the two $e_g$ orbitals are partially filled and hence active for the low-energy physics. Bridged by the apical oxygen $p_z$ orbital in the middle of the bilayer, the Ni-$3d_{3z^2-r^2}$ orbital in the top layer and that in the bottom layer have a large effective hopping and form a low-energy $\sigma$-bonding state and a high-energy anti-bonding state. This is absent for the Cu-$3d_{x^2-y^2}$ cuprates. The interplay between the two $e_g$ orbitals and the bilayer structure may lead to new electron correlation behaviors and different SC mechanisms.

In this Letter, we study the bilayer two-orbital model \cite{Luo2023arXiv} describing the La$_3$Ni$_2$O$_7$ under high pressure, including the most two relevant correlated orbitals, namely the Ni-$3d_{3z^2-r^2}$ ($z$) and $3d_{x^2-y^2}$ ($x$) orbitals. The model is solved by (cellular) dynamical mean-field theory (DMFT) \cite{Georges1996rmp,Maier2005RMP} at zero temperature. We analyze the relation between the two orbitals and find that they are orthogonal but entangled, which gives rise to concomitant two-orbital SC or Mott states. Orbital selective Mott (or SC) phases can emerge out when suppressing the non-local inter-orbital hoppings. When doping away from half filling, the ground state is a two-orbital $s_\pm$-wave SC, where the $z$-orbital SC is weaker than the $x$-orbital SC. To study the effect of the local inter-orbital spin correlation on the SC, we define an effective local inter-orbital spin coupling (LIOSC) split from the Hund's coupling \cite{Georges2013ARCMP}. We find that the LIOSC enhances significantly the electron correlations and share the $z$-orbital inter-layer antiferromagnetic correlation with the $x$ orbitals to help the $x$-orbital Cooper pairing.


\textit{Model.} We employ the bilayer two-orbital model proposed in Ref.~\cite{Luo2023arXiv}, which was obtained by the Wannier downfolding of the DFT band structure. Each layer is a square lattice. One layer is staked on the top of the other. The unit cell contains one site on the top layer and one site in the bottom layer. Each site contains both $e_g$ orbitals. The Hamiltonian is
\begin{equation}
\begin{aligned}
H&=H_0+H_I, \\
H_0&=\sum_{\rm{k} \sigma} \Psi_{\rm{k} \sigma}^{\dagger}  H(\rm{k}) \Psi_{\rm{k} \sigma}, \\
H_I&=U\sum_{il\alpha} (n_{il\alpha\uparrow}-\frac{1}{2})(n_{il\alpha\downarrow}-\frac{1}{2}) \\
&+U^{\prime}_{\rm{a}} \sum_{il\sigma}(n_{ilx\sigma}-\frac{1}{2})(n_{ilz\bar\sigma}-\frac{1}{2}) \\
&+U^{\prime}_{\rm{p}} \sum_{il\sigma}(n_{ilx\sigma}-\frac{1}{2})(n_{ilz\sigma}-\frac{1}{2}),
\end{aligned}\label{eq:h}
\end{equation}
where $\Psi_{\sigma} = (d_{Ax\sigma}, d_{Az\sigma}, d_{Bx\sigma}, d_{Bz\sigma})^{T}$ and $d_{s\sigma}$ annihilates an $s = Ax, Az, Bx, Bz$ electron with spin $\sigma$. $A$ and $B$ label the top layer and the bottom layer. $x$ and $z$ label $d_{x^2-y^2}$ and $d_{3z^2-r^2}$ orbitals, respectively. The lattice is shown in Fig.~2(a) of Ref.~\cite{Luo2023arXiv} while the model parameters are listed in Table~I of Ref.~\cite{Luo2023arXiv}. By convention, we set $U^{\prime}_{\rm{a}}$ = $U - 2J$, $U^{\prime}_{\rm{p}}$ = $U - 3J$, and $J = U/4$, where $J$ is the Hund's coupling. $U^{\prime}_{\rm{a}}$($U^{\prime}_{\rm{p}}$) is the inter-orbital repulsion strength for anti-parallel (parallel) spins. We choose $U = 4$ eV with eV being the energy unit. We further add a chemical potential for electron density tuning to the Halimtonian~(\ref{eq:h}). The band structure and Fermi surface of $H_{0}$ is shown in Fig.~3 of Ref.~\cite{Luo2023arXiv}.

Because of the mirror symmetry of the model, it is convenient to transform to the bonding ($+$) and anti-bonding ($-$) states for both $e_g$ orbitals $\Psi_{\pm \rm{k}\sigma} = (c_{x \pm \rm{k}\sigma}, c_{z \pm \rm{k}\sigma})^T$ with $c_{\alpha\pm \rm{k}\sigma} =  d_{\alpha \rm{k}A\sigma} \pm d_{\alpha \rm{k}B\sigma}$, in which the Hamiltonian becomes block-diagonal.

\textit{Method.} We solve this bilayer two-orbital model with (cellular) DMFT \cite{Georges1996rmp,Maier2005RMP} at zero temperature. The DMFT self-consistently maps the original model to a quantum impurity model. In this work, the impurity consists of two sites within a unit cell of the original lattice. So the local quantum fluctuation and inter-layer quantum correlation have been fully considered. By using the irreducible representation \cite{Koch2008PRB,Foley2019PRB}, our impurity model ensures the equivalence of the two layers. Local and inter-layer SC electron pairings between anti-parallel spins are both allowed, while in-plane non-local SC pairings, including $d$-wave pairings, are not allowed. We have calculated the bonding and anti-bonding SC pairing order parameters, 
\begin{equation}\label{eq:delta}
\Delta_{\alpha\pm} = \langle c_{\alpha\pm\uparrow}c_{\alpha\pm\downarrow}\rangle \end{equation}
as well as the local and inter-layer SC pairing order parameters, 
\begin{equation}\begin{aligned}
\Delta_{0 \alpha} & =\langle d_{A\alpha\uparrow}d_{A\alpha\downarrow} \rangle=\langle d_{B\alpha\uparrow}d_{B\alpha\downarrow} \rangle, \\
\Delta_{1 \alpha} & =\frac{1}{2}\langle d_{A\alpha\uparrow}d_{B\alpha\downarrow}+d_{B\alpha\uparrow}d_{A\alpha\downarrow}\rangle,
\end{aligned}\end{equation} 
where $\alpha = x, z$ labels different orbitals. Note that $\Delta_{\alpha\pm} = \Delta_{0\alpha} \pm \Delta_{1\alpha}$.

The electronic bath for the quantum impurity model is discretized into a number of bath sites. Then the quantum impurity model is solved by exact diagonalization (ED) \cite{Caffarel1994prl} and natural orbitals renormalization group (NORG) \cite{He2014prb,He2015prb} methods. The ED method can treat at most 8 bath sites and is used for most calculations. The NORG method can treat 12 bath sites so that the bath is better discribed, which is adopted for the calculation for Fig.~\ref{fig:3set}.


\textit{Orthogonality and entanglement between the two $e_g$ orbitals.} An electron can hop from a $z$ orbital of a Ni atom (say the atom at the origin) to an $x$ orbital of its nearby Ni atoms via $t^{xz}_3$ and $t^{xz}_4$, and further to other orbitals of other Ni atoms. But it will never hop to the $x$ orbitals of the original Ni atom and the other Ni atom (in the other layer) in the same unit cell and all the Ni atoms along the diagonal sites of the lattice. This is forbidden by the symmetry and quantum destructive interference. The all possible hopping paths will exactly and destructively interfere because, after a mirror operation against the $x=y$ plane, the $x$ oribitals change signs while the $z$ orbitals do not. As a result, on the $x=y$ plane (especially in a unit cell) there are no hybridization and Cooper pairing between the two $e_g$  orbitals, i.e., they are orthogonal to each other.

However, orthogonality does not mean independence. Fluctuations in the $z$ orbitals induce fluctuations in the $x$ orbitals, and vice versa, because of the nonlocal inter-orbital hoppings $t^{xz}_3$ and $t^{xz}_4$. As a result, the two $e_g$ orbitals are simultaneously SC or Mott insulating. One can not suppress SC in one orbital while leaving the other SC. And one orbital takes a Mott transition only when the other does simultaneously. In this aspect, the two $e_g$ orbitals entangle with each other.

\begin{figure}[htbp!]
  \includegraphics[width=8.6cm]{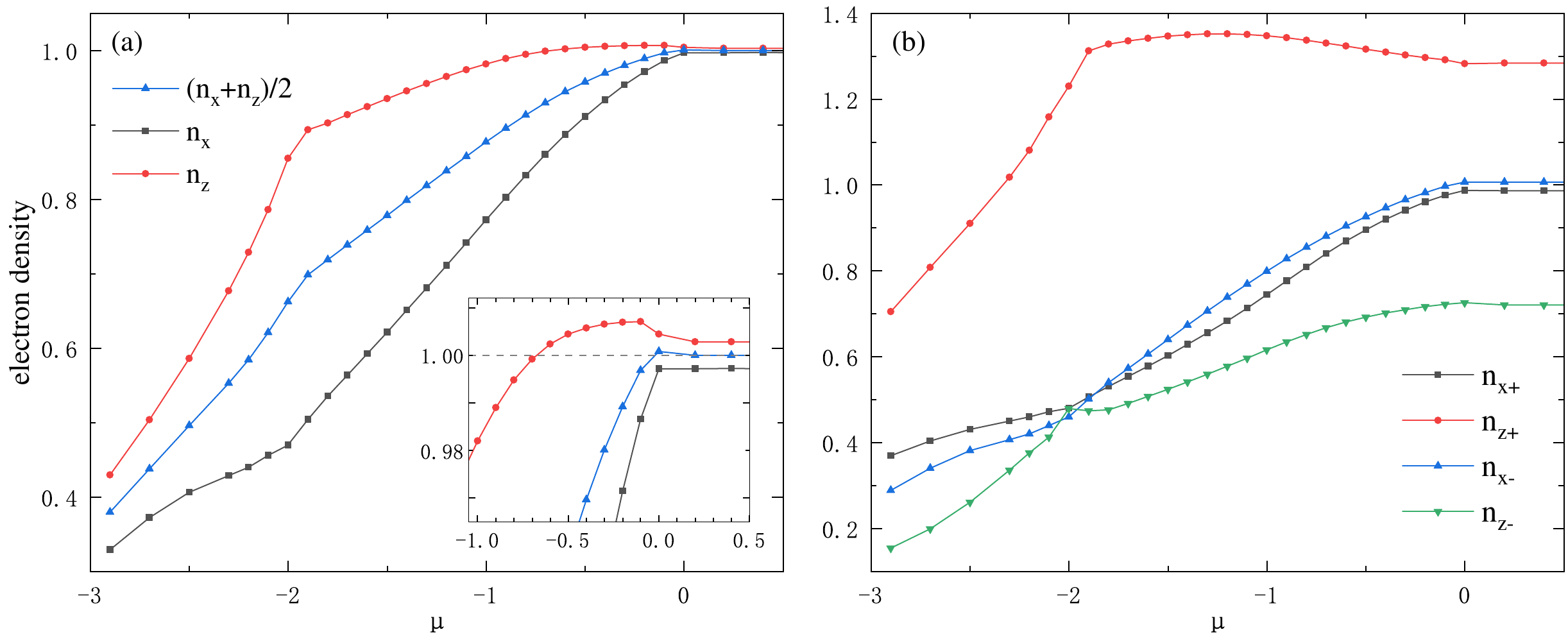}
  \caption{Electron density of a Ni atom as a function of the chemical potential $\mu$. (a) Electron densities for different orbitals and their average. Inset: enlarged view around half filling. When $\mu > -2$, the increase of $n_z$ becomes slowly. When $\mu$ exceeds $-0.7$, $n_z$ surpasses 1 without the $z$ orbitals being Mott insulating. As $n_x$ gradually approaches 1, $n_z$ drops back to 1, and the two orbitals simultaneously transition from SC to Mott insulating at $\mu = 0$. (b) Electron densities for the bonding and anti-bonding states of different orbitals. As $\mu$ decreases from 0, $n_{z+}$ increases counterintuitively and $n_{z-}$ is not small, implying that the $z$ orbitals are strongly correlated. A negative electronic compressibility is observed on the bonding state of $z$ orbital.}\label{fig:n-mu}
\end{figure}

\begin{figure}[htbp!]
  \includegraphics[width=8.6cm]{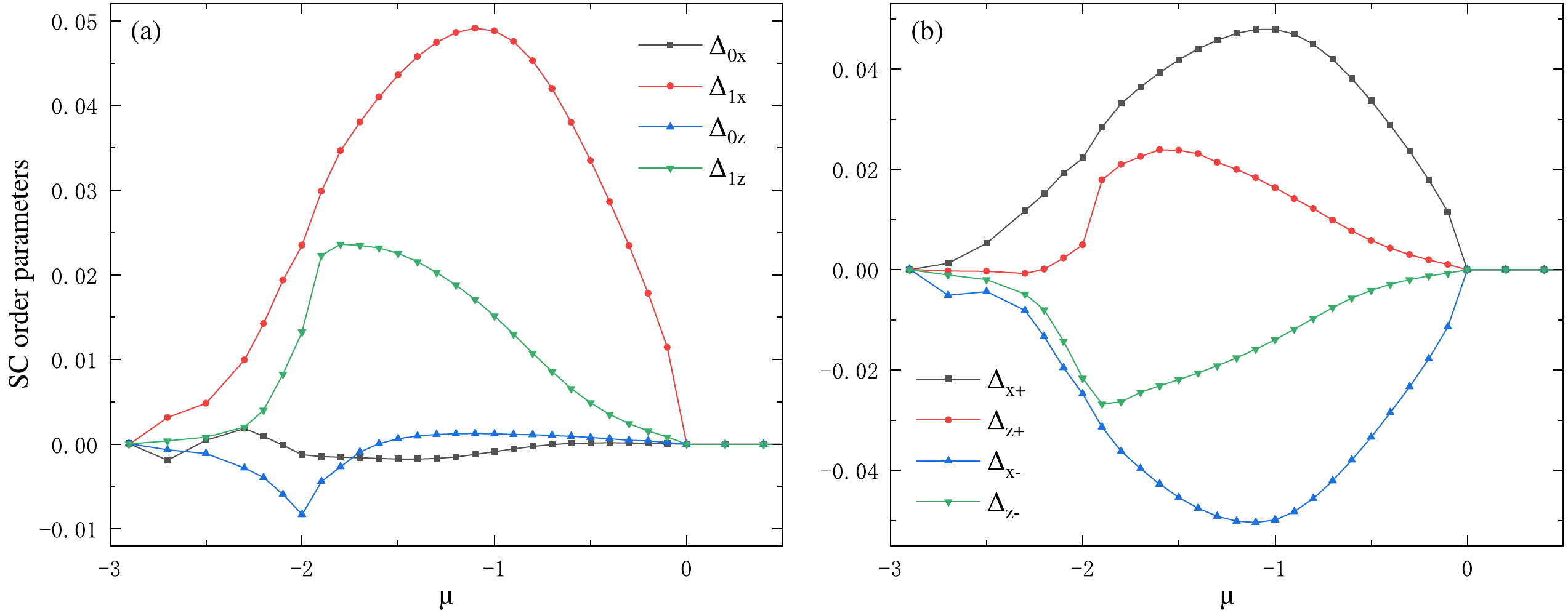}
  \caption{SC order parameters vs the chemical potential $\mu$. (a) Local ($\Delta_{0 \alpha}$) and inter-layer ($\Delta_{1 \alpha}$) pairings. In the whole SC regime, the inter-layer pairing is dominant, and $\Delta_{1x}$ is larger than $\Delta_{1z}$. (b) SC order parameters for the bonding ($\Delta_{\alpha+}$) and anti-bonding ($\Delta_{\alpha-}$) states. Both of the two orbitals are $s_\pm$-wave superconducting and become Mott insulating simultaneously at $\mu = 0$.}\label{fig:sc-mu}
\end{figure}

\textit{Simultaneous Mott transition of the two $e_g$ orbitals.} As we increase the chemical potential $\mu$, the electron densities of the two orbitals grow gradually. The electron density of $z$ orbitals ($n_z$) reaches half-filling first. But the $z$ orbitals do not encounter a Mott transition here because the $x$ orbitals are away from half-filling and can not be Mott insulating. As we further increase $\mu$, $n_z$ even surpasses 1 and then drops back to 1 as $n_x$ grows to 1, as shown in the inset of Fig.~\ref{fig:n-mu}(a). Then the two orbitals take a Mott transition simultaneously (Fig.~\ref{fig:n-mu}), the SC disappears (Fig.~\ref{fig:sc-mu}), and both the electron densities become constant (Fig.~\ref{fig:n-mu}(a)). In other words, there are no orbital selective Mott transition (OSMT) in the two $e_g$ orbitals.

\textit{Concomitant $s_\pm$-wave SCs.} When doping away from half filling, the ground state of the system becomes superconducting. The electrons pair mainly between the two $z$ orbitals (as well as the two $x$ orbitals) of the two Ni atoms in a unit cell, namely $\Delta_{1z} \neq 0$ and $\Delta_{1x} \neq 0$, while $\Delta_{0z}$ and $\Delta_{0x}$ are small (Fig.~\ref{fig:sc-mu}(a)). Because of the orthogonality between the $x$ orbitals and $z$ orbitals in a unit cell, there are no electron pairings between them. As a result, $\Delta_{z\pm} = \Delta_{0z} \pm \Delta_{1z} \ne 0$ and $\Delta_{x\pm} = \Delta_{0x} \pm \Delta_{1x} \ne 0$ (Fig.~\ref{fig:sc-mu}(b)) but $\langle c_{x\pm\uparrow}c_{z\pm\downarrow}\rangle = 0$, i.e., both of the two $e_g$ orbitals are $s_\pm$-wave pairing, but there are no Cooper pairing between them. Nevertheless, we want to emphasize that, because of the entanglement between the two orbitals, the two $s_\pm$-wave SCs are concomitant and there are no orbital selective SC (OSSC) in the two $e_g$ orbitals.

\begin{figure}[htbp!]
  \includegraphics[width=8.6cm]{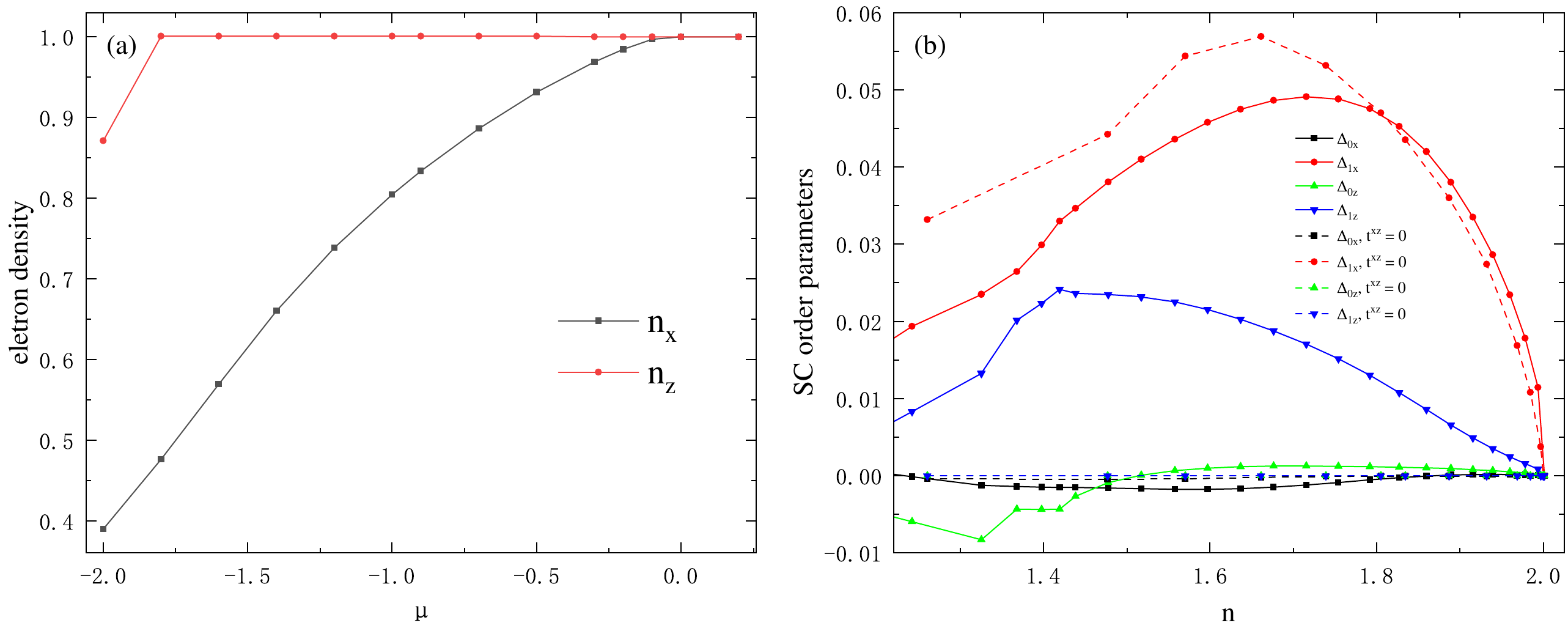}
  \caption{(a) Electron densities for different orbitals and (b) SC order parameters as functions of chemical potential $\mu$ after $t^{xz}_3$ and $t^{xz}_4$ being suppressed to zero. Solid lines represent order parameters for normal hopping parameters, and dashed lines for hopping parameters with $t^{xz}_3$ and $t^{xz}_4$ suppressed to zero. An orbital selective Mott (or SC) phase has been observed after suppressing the nonlocal inter-orbital hopping parameters $t^{xz}$. All SC order parameters have been suppressed except $\Delta_{1x}$.}\label{fig:notxz}
\end{figure}

To show the crucial role of the nonlocal inter-orbital hoppings for the simultaneous Mott transition and the concomitant $s_\pm$-wave SCs, we do another calculation with them turned off, i.e. setting $t^{xz}_3 = t^{xz}_4 = 0$. In this case, the two $e_g$ orbitals couple to each other only through the two-body interactions, the inter-orbital hybridization disappears, and the inter-orbital charge flow is forbidden. Consequently, we have observed an OSMT and OSSC (Fig.~\ref{fig:notxz}), where, as the chemical potential increases, the $z$ orbitals reach half filling first and become Mott insulating and their SC has been suppressed. In contrast, the $x$-orbital SC is almost unaffected, which is robust and does not rely on the SC of the $z$ orbitals.

\begin{figure}[htbp!]
  \includegraphics[width=8.6cm]{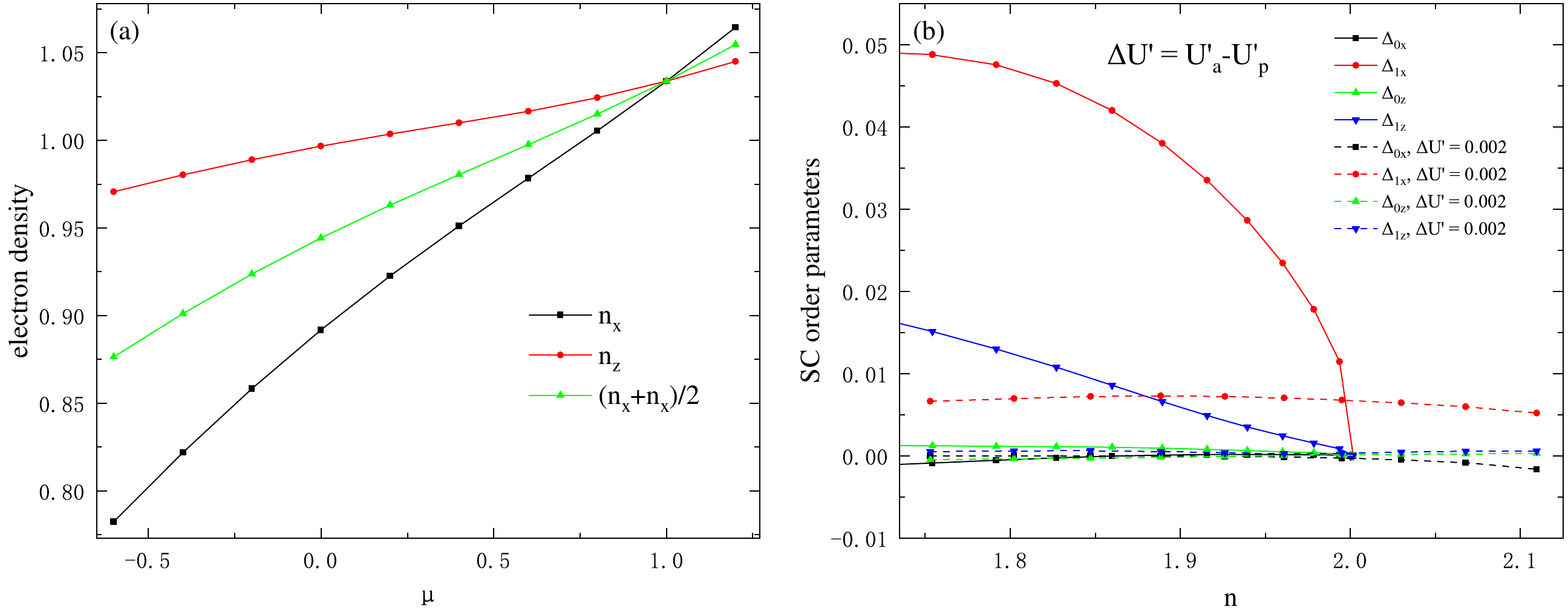}
  \caption{(a) Electron densities and (b) SC order parameters with $\Delta U^{\prime} = U^{\prime}_{\rm{a}}-U^{\prime}_{\rm{p}}=0.002$ and $U^{\prime}_{\rm{a}}+U^{\prime}_{\rm{p}}$ unchanged.  An almost linear relation for both $n_x$ and $n_z$ as $\mu$ increasing. In (b), solid lines represent order parameters for $\Delta U^{\prime} = 1$ and dashed lines for $\Delta U^{\prime} = 0.002$. $\Delta_{1x}$ has been suppressed to a small magnitude, and other SC order parameters almost disappear.}\label{fig:3set}
\end{figure}

\textit{Dramatically enhanced electron correlations and the effect on the SC by the local inter-orbital spin coupling.} Strong correlation, especially antiferromagnetic (AFM) spin correlation, induced by the Coulomb interaction is believed to play a key role in unconventional SCs. In La$_3$Ni$_2$O$_7$ under high pressure, a strong inter-layer AFM correlation between the two $z$ orbitals can be induced by the intra-orbital Hubbard interaction and the effective hopping between the two $z$ orbitals mediated by the oxygen-$2p_z$ orbital between them. In contrast, this mechanism is missing for the $x$-orbital conterpart. But the $x$-orbital $s_\pm$-wave SC is robust as shown above. A sound explanation for its AFM correlation is that the $x$ orbitals share the AFM correlation of the $z$ orbitals through the Hund's coupling, which tends to align the spins of the two $e_g$ orbitals in a single Ni atom. We study this point in more detail.

To this end, we define a local inter-orbital spin coupling (LIOSC) $\Delta U^\prime = U^\prime_{\rm a} - U^\prime_{\rm p}$, which exactly accounts for the spin correlation between the two $e_g$ orbitals in a Ni atom. We do a calculation with reducing $\Delta U^\prime$ from its normal value $J$ to a negligible value $0.002J$ while keeping $(U^\prime_{\rm a} + U^\prime_{\rm p})/2$ unchanged. Surprisingly, the previous Mott insulating state with a large Mott gap disappears (Fig.~\ref{fig:3set}(a)), and the SCs of the two orbitals are greatly suppressed (about 7 times weaker for both the two orbitals), showing that the LIOSC $\Delta U^\prime$ can dramatically enhance the electron correlations and is crucial to the high $T_c$ SC for this two-orbital system. This supports the above mechanism.

\begin{figure}[htbp!]
  \includegraphics[width=8.6cm]{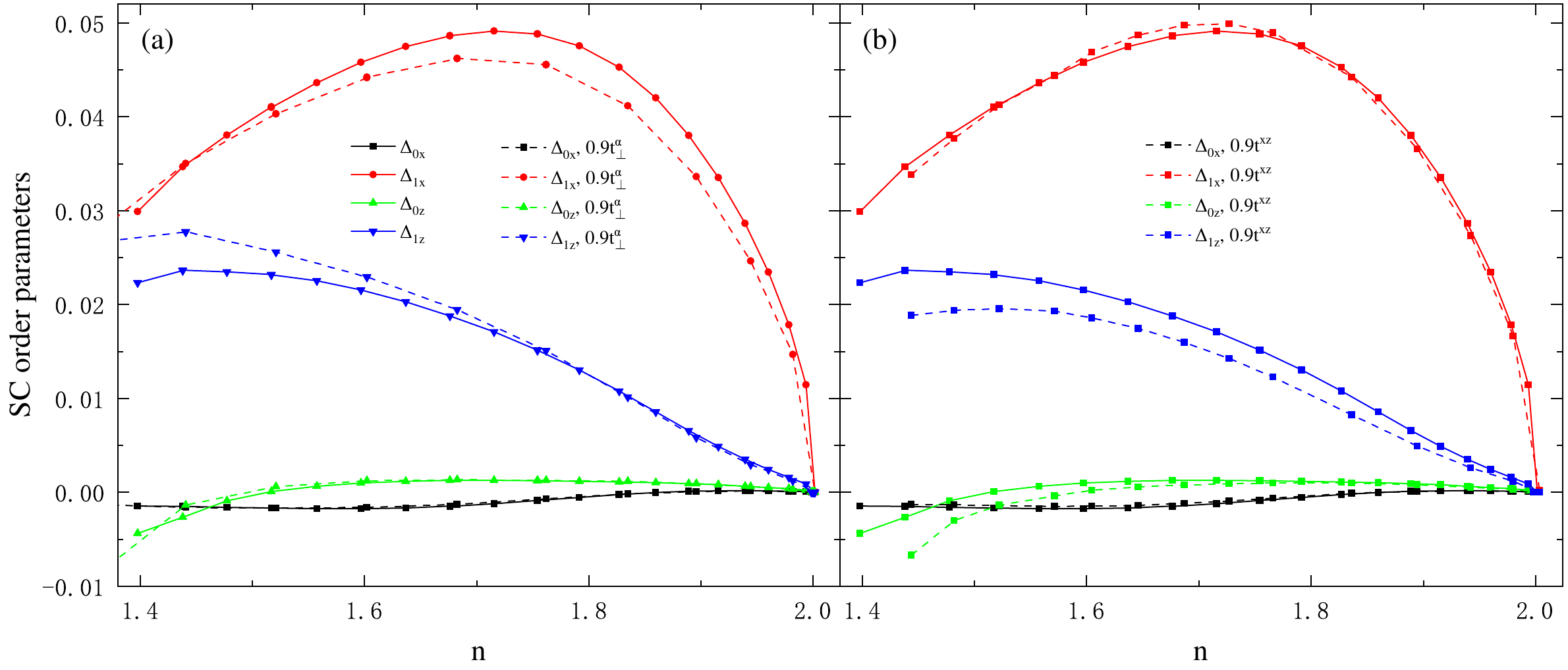}
  \caption{SC order parameters vs electron density for different values of hopping parameters. Solid lines represent order parameters for normal hopping parameters, and dashed lines for hopping parameters which are suppressed. (a) $t_{\perp}^{\alpha}$ for both orbitals are reduced by 10\%. (b) $t^{xz}_3$ and $t^{xz}_4$ are reduced by 10\%.}\label{fig:tpp-txz}
\end{figure}

On the other hand, the inter-layer AFM correlation depends directly on the inter-layer hoppings $t_\perp^z$ and $t_\perp^x$. Reducing them suppresses the $x$-orbital SC (Fig.~\ref{fig:tpp-txz}(a)), implying that the $x$-orbital SC relies on the inter-layer AFM correlation of the $z$ orbitals. This also supports the above picture.

Although the inter-layer AFM correlation for the $z$ orbitals is strong, its SC is weaker than the $x$-orbital SC. This is because the coherent moving of the $z$ orbital Cooper pairs is more difficult since all the in-plane hoppings for the $z$ orbitals are small. When reducing $t_3^{xz}$ and $t_4^{xz}$, the $z$-orbital SC is suppressed while the $x$-orbital SC is not (Fig.~\ref{fig:tpp-txz}(b)).

\textit{Summary.} To study the correlation effects and the high-$T_c$ SC in La$_3$Ni$_2$O$_7$ under high pressure, we have adopted the (cellular) DMFT on a bilayer two-orbital Hubbard model. The two active $e_g$ orbitals and their inter-layer correlations are included. We have observed concomitant two-orbital $s_\pm$-wave SC. The two orbitals are orthogonal to each other because of symmetry and quantum destructive interference. Meanwhile, they entangle with each other because of the nonlocal inter-orbital hoppings. As a result, each orbital forms Cooper pairs by itself and the two orbitals become SC or Mott insulating simultaneously when tuning the electron density. The $z$ orbitals form $s_\pm$-wave SC because of their strong inter-layer hoppings and antiferromagnetic correlation. But its SC order parameter is not large because of their weak in-plane hoppings. The $x$ orbitals form strong $s_\pm$-wave SC by sharing the inter-layer antiferromagnetic correlation from that of the $z$ orbitals. This sharing is accounted for by the local inter-orbital spin correlation caused by the effective local inter-orbital spin coupling (LIOSC). When suppressing the LIOSC, the Mott state disappears and the SC is suppressed remarkably, implying that the LIOSC enhances pronouncedly the electron correlation and is crucial to the SC. When suppressing the nonlocal inter-orbital hoppings, the two orbitals do not entangle with each other anymore and an orbital selective Mott or SC phase has been observed.

\begin{acknowledgments}
  This work was supported by National Natural Science Foundation of China (Grant No. 11934020). Computational resources were provided by Physical Laboratory of High Performance Computing in RUC.
\end{acknowledgments}

\bibliography{lnosc}

\end{document}